\documentclass[11pt]{article}
\usepackage[utf8]{inputenc}
\usepackage{geometry}
\usepackage{amsmath}
\usepackage[dvipsnames]{xcolor}
\usepackage{dsfont}
\usepackage{setspace}
\usepackage{changepage}
\usepackage[breakable]{tcolorbox}
\usepackage{dirtytalk}
\usepackage{algorithm}
\usepackage{algpseudocode}
\usepackage{graphicx}
\usepackage{capt-of}
\usepackage[symbol]{footmisc}
 
\DeclareMathOperator*{\argmin}{arg\min}

\usepackage{authblk}
\usepackage{hyperref}
\hypersetup{
    colorlinks=true,
    linkcolor=black,
    filecolor=magenta,      
    urlcolor=cyan,
}

\title{\fontseries{b}\selectfont Self-STORM: Deep Unrolled Self-Supervised Learning \\ for Super-Resolution Microscopy}
\author[1]{Yair Ben Sahel}
\author[1]{Yonina C. Eldar}
\affil[1]{\small Department of Computer Science and Applied Mathematics, Weizmann Institute of Science, Israel}
\date{}

\begin{document}

\maketitle

\begin{adjustwidth}{30pt}{30pt}
\small{The use of fluorescent molecules to create long sequences of low-density, diffraction-limited images enables highly-precise molecule localization. However, this methodology requires lengthy imaging times, which limits the ability to view dynamic interactions of live cells on short time scales. Many techniques have been developed to reduce the number of frames needed for localization, from classic iterative optimization to deep neural networks. Particularly, deep algorithm unrolling utilizes both the structure of iterative sparse recovery algorithms and the performance gains of supervised deep learning. However, the robustness of this approach is highly dependant on having sufficient training data. In this paper we introduce deep unrolled self-supervised learning, which alleviates the need for such data by training a sequence-specific, model-based autoencoder that learns only from given measurements. Our proposed method exceeds the performance of its supervised counterparts, thus allowing for robust, dynamic imaging well below the diffraction limit without any labeled training samples. Furthermore, the suggested model-based autoencoder scheme can be utilized to enhance generalization in any sparse recovery framework, without the need for external training data.}
\end{adjustwidth}


\section{Introduction}
The resolution limit of optical imaging systems was long considered to be determined by Abbe's diffraction limit, which is hundreds of nanometers at best for modern light microscopes. When dealing with labeled samples, like in fluorescence microscopy, one may overcome the diffraction limit by distinguishing between the photons coming from two neighboring fluorophores \cite{sub_diff}. One way to distinguish neighboring molecules, is by utilizing photo-activated or photo-switching fluorophores to separate fluorescent emission in time; this is the basis for Single Molecule Localization Microscopy (SMLM) techniques such as Photo-Activated Localization Microscopy (PALM) and Stochastic Optical Reconstruction Microscopy (STORM) \cite{SMLM1, SMLM2}. These methods take a sequence of diffraction-limited images, produced by a sparse set of emitting fluorophores with minimally overlapping point-spread functions (PSFs). This allows for the emitters to be localized with high precision by relatively simple post-processing. This enables imaging sub-cellular features and organelles within biological cells with unprecedented resolution. The low emitter density concept requires lengthy imaging times to achieve full coverage of the imaged specimen on the one hand, and minimal overlap between PSFs on the other. Thus, this concept in its classical form has low temporal resolution, limiting its application to slow-changing specimens and precluding more general live-cell imaging.

To circumvent the long acquisition periods required for SMLM methods, a variety of techniques have emerged, which enable the use of a smaller number of frames for reconstructing the 2-D super-resolved image \cite{falcon, CSSTORM, SOFI, SPARCOM, SPARCOM_MATH, annapalm, deepSTORM, decode}. These techniques take advantage of prior information regarding either the optical setup, the geometry of the sample, or the statistics of the emitters. In \cite{falcon, CSSTORM}, molecule localization is performed via frame-by-frame recovery using sparse coding techniques \cite{sampling, CS}. Super-resolution optical fluctuation imaging (SOFI) employs high-order statistical analysis (cumulants) of temporal fluctuations \cite{SOFI} to enhance spatial resolution by reducing the effective point spread function size based on the square root of the cumulant order. However, the use of statistical orders higher than two is limited due to signal-to-noise ratio (SNR), dynamic range expansion, and temporal resolution considerations, collectively resulting in considerably lower spatial resolution than that achieved by PALM and STORM techniques. Solomon et al. suggested combining sparse recovery principles with SOFI, resulting in a sparsity-based approach for super-resolution microscopy from correlation information of high emitter-density frames, dubbed SPARCOM \cite{SPARCOM, SPARCOM_MATH}. SPARCOM capitalizes on sparsity within the correlation domain while assuming that the blinking emitters are uncorrelated across both time and space. The use of SPARCOM was shown to increase the number of detected emitter locations when compared to sparse recovery performed directly on the signal itself \cite{SR_CORR}, leading to notable enhancements in both temporal and spatial resolution when compared to its counterparts. However, SPARCOM requires adjustment of optimization parameters and explicit knowledge of the impulse response of the imaging system. It is also computationally expensive and converges slowly.

Deep learning approaches have overcome some of these disadvantages \cite{deepSTORM, decode, anna-palm, ozcan}. One such approach is Deep-STORM \cite{deepSTORM}, which takes high-emitter-density frames as inputs and employs a U-net architecture to reconstruct a super-resolved image. It leverages prior knowledge about the optical setup by training on a dataset generated through simulation with the matching optical parameters. However, as demonstrated in the results section, Deep-STORM's performance deteriorates when applied to test data with dissimilar imaging parameters compared to those seen during training. Additionally, Deep-STORM has many trainable parameters and lacks an easily interpretable structure, which could be particularly valuable in biological contexts. An alternative deep learning technique is DECODE (deep context dependant) \cite{decode}, which uses information from multiple consecutive frames to predict the probability of detection, along with the sub-pixel localization and localization uncertainty of each detected emitter. Similar to Deep-STORM, it is also trained via simulator learning: the training set is comprised of random ground-truth (GT) emitter coordinates and matching synthetic images simulated from a forward model of the image formation process. DECODE allows for 20-fold improvement in prediction speed and reduces the localization error up to twofold compared to Deep-STORM, thanks to the use of temporal context and continuous sub-pixel coordinates instead of super-resolved voxels. However, as in the case of Deep-STORM, it relies on an accurate PSF model and proper parameters. In cases where there are substantial differences in imaging parameters, artifacts dominate the model's predictions, as shown in the results section.

The limitations of learning-based techniques can be overcome by taking advantage of both the interpretability of iterative techniques and the flexibility of deep learning, via algorithm unrolling \cite{yair, boyd}. Algorithm unrolling replaces the iterations of iterative algorithms with neural networks which perform the same mathematical operation. By doing so, parameters which would have to be specified explicitly or tuned empirically are learned automatically, and relevant context ignored by the algorithm may be incorporated into the learned model. Gregor and LeCun \cite{LISTA} applied the unrolling framework to the Iterative Shrinkage and Thresholding Algorithm (ISTA), resulting in Learned ISTA (LISTA). For a given number of iterations/layers, the trained LISTA network obtains lower prediction error than ISTA, and achieves faster convergence. Dardikman-Yoffe et al. have recently incorporated LISTA into SPARCOM, resulting in a method called Learned SPARCOM (LSPARCOM) \cite{LSPARCOM}, which outperforms classical algorithms such as SPARCOM and generalizes better than learning-based methods like Deep-STORM. This was facilitated by the use of a compact neural network which utilizes a model-based framework, thus eliminating the need for data-specific training both in terms of structure and imaging parameters. However, similar to previous learning-based methods, LSPARCOM is trained using supervised learning from simulated data. Thus, as shown in the results section, recovery performance degrades when the data used to train the model is generated differently from that being analyzed, leaving the challenge of generalization yet to be resolved.

We address this challenge by utilizing self-supervised learning. Self-supervised image super-resolution methods \cite{dip, zssr, DualBSR, smore, sisr_rev} enhance the resolution of images without relying on external high-resolution reference images or paired training data. Instead, they leverage the information present within the same low-resolution image to guide the up-scaling process. One of these methods is Zero-Shot Super-Resolution (ZSSR) \cite{zssr}, which exploits the internal recurrence of information within a single image by training a compact, image-specific model during testing, using examples exclusively extracted from the provided low-resolution input image. As a result, the trained model is adapted to the specific settings of the input image. This alleviates the need for prior training data and allows for super-resolution on images where the image formation process is unknown and non-ideal. However, it still requires a rough estimate of the down-scaling kernel, and its generic model architecture does not adapt well to the SMLM setting, as demonstrated in the results section.

In this paper, we introduce Self-STORM: a new approach to analyze SMLM data, which incorporates self-supervised learning into the unrolling framework. Self-STORM integrates the previously-supervised unrolled model (i.e., LISTA) into a model-based autoencoder, which learns only from low-resolution measurements. This relieves the need for external training samples, resulting in a learned model that generalizes well, is interpretable, and requires only a small number of parameters, without relying on explicit knowledge of the optical setup or requiring fine-tuning of optimization parameters. We compare Self-STORM against formerly known methods for SMLM image reconstruction: the classical iterative algorithm, SPARCOM \cite{SPARCOM}, leading deep-learning based localization methods for high emitter-density frames, Deep-STORM \cite{deepSTORM} and DECODE \cite{decode}, the unrolled supervised model, LSPARCOM \cite{LSPARCOM}, and a self-supervised technique, based on ZSSR \cite{zssr}. We do so by analyzing their performance on both simulated and experimental data. The results show that Self-STORM yields results that are on par with those obtained by other supervised techniques, on data that is similar to their training sets. The results show that Self-STORM outperforms any other method for data that is substantially different than the data it was trained on. Moreover, Self-STORM yields results that are on par with those obtained by other supervised techniques, on data that is similar to their training datasets. In addition, we show that Self-STORM produces excellent reconstructions for SMLM data with ultra-high emitter densities, thus paving the way to fast dynamic live-cell imaging. This is made possible by the use of a parameter-efficient neural network which utilizes a self-supervised model-based framework, thus not requiring labeled training data of any kind. Furthermore, runtime comparison shows that even though our network is trained at test time, its \textit{train and test} runtime is comparable to the \textit{test} runtime of any other method. Thus, Self-STORM paves the way to true live-cell SMLM imaging using a compact, interpretable deep network that requires no pre-training, can generalize well to any setting, and requires a small number of frames over time. Furthermore, the self-supervised model-based framework has the potential to be applied across various sparse recovery scenarios, offering an effective means to improve generalization without the need for pre-existing training data. This versatile approach can be harnessed to enhance performance in diverse contexts where sparse recovery is a fundamental task, obviating the necessity for prior data-dependent training.

The rest of the paper is organized as follows. In Section 2 we formulate the super-resolution localization task as a sparse recovery problem. In Section 3, we describe iterative sparse recovery and its limitations, and how algorithmic unrolling may be adapted to address some of its shortcomings. Section 4 details the proposed integration of self-supervised learning into the unrolling framework, resulting in Self-STORM. In Section 5, we provide representative results of Self-STORM, compared to other known techniques for SMLM data analysis. Finally, we discuss both the impact and limitations of our approach in Section 6.

Throughout the paper, $t$ represents time, $x$ represents a scalar, $\mathbf{x}$ denotes a vector, $\mathbf{X}$ denotes a matrix, and $\mathbf{I}_{N}$ is the $N \times N$ identity matrix. Subscript $x_l$ is the $l$’th element of $\mathbf{x}$, and $\mathbf{x}_l$ is the $l$’th column of $\mathbf{X}$. Superscript $\mathbf{x}^{(k)}$ represents $\mathbf{x}$ at iteration $k$, and $\mathbf{A}^T$ is the transpose of $\mathbf{A}$.


\section{Problem Formulation}
In the SMLM setting, we aim to recover a single $N \times N$ high-resolution image, corresponding to the locations of the emitters on a fine grid, from a set of $T$ low-resolution $M \times M$ frames, with $N>M$. For each frame $i \in \{1, 2, ... ,T\}$, we consider the field-of-view (FOV) as a high-resolution grid which is vectorized to form $\mathbf{x}_i\in\mathds{R}^{N^2}$. The locations of emitters in the sample are modeled by assigning each element of $\mathbf{x}_i$ a value related to the number of photons emitted from that location within the FOV. Given that the imaging sensor has $M \times M$ pixels (with $N > M$), we model the imaging process as multiplication by a matrix $\mathbf{A}\in\mathds{R}^{M^2 \times N^2}$, in which element $(i,j)$ is the proportion of signal emitted from location $j$ on the high-resolution grid that will be detected at pixel $i$ of the sensor. Thus defined, the columns of $\mathbf{A}$ represent the PSF of the imaging system, such that column $j$ of $\mathbf{A}$ is the PSF of the system for a point source at location $j$. The (vectorized) measured image frame is then $\mathbf{y}_i = \mathbf{Ax}_i$, with $\mathbf{y}\in\mathds{R}^{M^2}$. Note that the non-zero entries in each $\mathbf{x}_i$ correspond to the locations of activated emitters. Thus, given $\mathbf{y}_1, ... , \mathbf{y}_T$ and $\mathbf{A}$, we aim to recover $\mathbf{x}_1, ... , \mathbf{x}_T$.The support of the sum over $\mathbf{x}_i$ yields all emitters' locations on the high-resolution grid, which we reshape to our desired $N \times N$ image.

This inference problem can be formulated as a least-squares optimization problem: 
\begin{equation}
\label{eqn:ls}
\mathbf{\hat{x}}_i = \argmin_{\mathbf{x}}\|\mathbf{y}_i-\mathbf{Ax}\|^{2}_{2}.
\end{equation}
Even if $\mathbf{A}$ is known perfectly, as long as $N>M$, $\mathbf{A}$ will have a nontrivial null space, so that the optimization problem is under-determined. Leveraging knowledge of the biological structure of $\mathbf{x}$ can resolve this issue. Specifically for SMLM, we know that each frame contains a sparse set of light-emitting molecules. Let $K$ be the number of emitters on the $N \times N$ grid. The sparsity of emitters in each frame implies that $K << N$, and that every column $\mathbf{x}_i$ of $\mathbf{X}$ is at least $K$-sparse. This allows us to choose a sparse optimization technique \cite{sampling}\cite{CS}, such as the well-known LASSO \cite{LASSO}, to recover $\mathbf{x}_i$:

\begin{equation}
\label{eqn:sls}
\mathbf{\hat{x}}_i = \min_{\mathbf{x}_i}\|\mathbf{y}_i-\mathbf{Ax}_i\|^{2}_{2}+\lambda\|\mathbf{x}_i\|_{1}.
\end{equation}

In particular, by correctly tuning $\lambda$, the support of the sum over minimizers $\{\mathbf{\hat{x}}_i\}_{i=i...T}$ provides the locations of all emitters. Since we wish to circumvent long acquisition periods, $T$ should be as small as possible (i.e., minimal number of frames per super-resolved image), such that we can successfully recover all emitters' locations.


\section{Algorithm Unrolling for Super-Resolution Microscopy}
Once a problem is framed as a sparse optimization of the form (\ref{eqn:sls}), one may use an iterative algorithm such as ISTA to find $\mathbf{\hat{x}}_i$, as described in Algorithm \ref{alg:ISTA}. Given (\ref{eqn:sls}), ISTA estimates $\mathbf{x}$, taking as inputs the measurement matrix $\mathbf{A}$, the measurement vector $\mathbf{y}_i$, the regularization parameter $\lambda$, and $L$, a Lipschitz constant of $\nabla \|\mathbf{Ax}_i-\mathbf{y}_i\|_{2}^{2}$.

\begin{algorithm}[H]
\setstretch{1}
\caption{ISTA}
\label{alg:ISTA}
\begin{algorithmic}[1]
\Require{$\mathbf{y}_i$, $\mathbf{A}$, $\lambda$, $L$, number of iterations $k_{max}$ }
\Ensure{$\mathbf{\hat{x}}_i$}
\State $\mathbf{\hat{x}^{(1)}}_i = 0$, $k = 1$
\While{$k<k_{max}$}
\State $\mathbf{\hat{x}^{(k+1)}}_i = \mathbf{\mathcal{T}}_{\frac{\lambda}{L}}(\mathbf{\hat{x}^{(k)}}_i-2L\mathbf{A^{T}}(\mathbf{A}\mathbf{\hat{x}^{(k)}}_i-\mathbf{y}_i))$
\State $k\leftarrow k+1$
\EndWhile
\State $\mathbf{\hat{x}}_i = \mathbf{\hat{x}^{(k_{max})}}_i$
\end{algorithmic}
\end{algorithm}

\noindent The operator $\mathbf{\mathcal{T}}_{\frac{\lambda}{L}}(\cdot)$ in Algorithm \ref{alg:ISTA} is the positive soft thresholding operation, which is equal to the
shifted rectified linear unit (ReLU), defined by:

\begin{equation}
\label{eqn:th_op}
    \mathbf{\mathcal{T}}_{\frac{\lambda}{L}}(x) =ReLU\left(x-\frac{\lambda}{L}\right) = \max\left\{x-\frac{\lambda}{L},0\right\},
\end{equation}

where $x$ is scalar. When applied to vectors and matrices, $\mathbf{\mathcal{T}}_{\frac{\lambda}{L}}$ operates element-wise. The use of the soft thresholding operator is derived from (\ref{eqn:sls}), as it is the proximal operator of the $L_1$-norm. The elements of $\mathbf{x}_i$ represent the intensity of the emitters, and therefore are necessarily non-negative. Thus, we use a positive soft thresholding operator, rather than a standard soft thresholding operator (after which the original ISTA is named). Note that the argument of $\mathbf{\mathcal{T}}_{\frac{\lambda}{L}}(\cdot)$ can be rewritten as the sum of matrix-vector products with $\mathbf{y}$ and $\mathbf{\hat{x}^{(k)}}_i$:
\begin{equation}
\begin{gathered}
\label{eqn:ISTA_step}
    \mathbf{\hat{x}^{(k)}}_i-2L\mathbf{A}^{T}(\mathbf{A}\mathbf{\hat{x}^{(k)}}_i-\mathbf{y}_i) = \mathbf{W_{0}}\mathbf{y}_i + (\mathbf{I-W})\mathbf{\hat{x}^{(k)}}_i, 
\end{gathered}
\end{equation}
where $\mathbf{W_{0}} = 2L\mathbf{A}^{T}$ and $ \mathbf{W} = 2L\mathbf{A}^{T}\mathbf{A}$.

The iterative approach described above requires both prior knowledge of the PSF of the optical setup for the calculation of the measurement matrix, which is not always available, and a wise choice of regularization factor $\lambda$, which is generally performed heuristically. In order to overcome these shortcomings, we may apply algorithm unrolling and use LISTA instead of ISTA. The idea at the core of deep algorithm unfolding, as first suggested by Gregor and LeCun \cite{LISTA}, is using the algorithmic framework to gain interpretability and domain knowledge, while inferring optimal parameters from the data itself. In this strategy, the design of a neural network architecture is tailored to the specific problem, based on a well-founded iterative mathematical formulation for solving the problem.

In the case of ISTA, (\ref{eqn:th_op}) and (\ref{eqn:ISTA_step}) imply that the iterative step of the algorithm can be modeled by the sum of fully-connected layers and an activation function with learned threshold. Moreover, in our setting, we may replace the multiplication by $\mathbf{W_{0}}$ and multiplication by $\mathbf{W}$ in (\ref{eqn:ISTA_step}) with convolutions: Recall that each column $j$ of $\mathbf{A}$ is an $M^2$ long vector representing the PSF of the system for a point source at location $j$. Thus, each row $j$ in $\mathbf{W_{0}} = 2L\mathbf{A}^{T}$ corresponds to the PSF of the system for a point source at location $j$ (multiplied by a constant). Therefore, the first multiplication operation in (\ref{eqn:ISTA_step}) is equivalent to convolving the low-resolution input $\mathbf{y}_i$ with the system's PSF, using sub-pixel strides of size $\frac{M}{N}$, creating an up-sampling effect. Similarly, we note that $\mathbf{A}^T\mathbf{A}$ is a symmetric $N^2 \times N^2$ matrix, where each column (and row) $j$ is a $N^2$ long vector corresponding to a specific kernel, shifted to location $j$. This kernel is composed of the sum of the PSF from each point on the high-resolution grid, shifted and weighted according to the PSF from every other point on the high-resolution grid. This implies that the second multiplication operation in (\ref{eqn:ISTA_step}) is equivalent to convolving $\mathbf{\hat{x}^{(k)}}_i$ with said kernel.

To increase sparsity, we replace the positive soft thresholding operator with a differentiable, sigmoid-based approximation of the positive hard-thresholding operator \cite{l0-relu}, which effectively changes the $L_1$ regularization in (\ref{eqn:sls}) to $L_0$ regularization \cite{LSPARCOM}. This smooth activation function, denoted as $S^{+}_{\alpha, \beta}(\cdot)$, has two trainable parameters $(\alpha, \beta)$:
\begin{equation}
\begin{gathered}
\label{eqn:activ}
    S^{+}_{\alpha, \beta}(x) = \frac{ReLU(x)}{1+exp(-\beta(|x|-\alpha)},
\end{gathered}
\end{equation}
where $x$ is scalar. When applied to vectors and matrices, $S^{+}_{\alpha, \beta}(\cdot)$ operates element-wise. Since different frames typically have different emitter intensities, e.g. due to heterogeneous illumination, we adopt the variation of $S^{+}_{\alpha, \beta}(\cdot)$ offered in \cite{LSPARCOM}, and define the trainable values of the smooth activation function to be relative rather than absolute, denoted as $\alpha_0$ and $\beta_0$:
\begin{equation}
\begin{gathered}
\label{eqn:activ_params}
   \alpha = i_1 + (i_{99} - i_1)\alpha_0, \\ \beta = \frac{\beta_0}{\alpha},
\end{gathered}
\end{equation}
where $0 \leq \alpha_0 \leq 1$ and $i_1, i_{99}$ are the first and 99-th percentiles of the input to the activation layer. This enables applying a specific threshold per each input frame, as opposed to a global threshold across all frames.

Finally, we concatenate these convolution and activation layers together, resulting in a deep neural network (i.e. LISTA), that has the same form as the operation performed by running ISTA over multiple iterations. The LISTA network may be optimized using supervised learning, with training data consisting of paired examples of the locations $\mathbf{x}_i$ and measurements $\mathbf{y}_i$ from (\ref{eqn:sls}). Training data may be obtained, for example, from measurement simulations with known ground truth locations \cite{deepSTORM, decode}. As a result, we get highly accurate results when test data is similar to the training data; this is shown in Figure \ref{fig:BTHD}, where the best-performing method is the supervised, LISTA-based LSPARCOM, whose training set had identical structure to that of the test set. On the other hand, reliance on labelled training examples leads to poor localization accuracy for data that is significantly different than the training set. This is shown in Figure \ref{fig:MTO}, where supervised algorithms yield inaccurate results for data that is different from their respective training sets. Therefore, we propose Self-STORM: a self-supervised learning scheme, which enhances robustness by alleviating the need for labeled input-output examples.


\section{Deep Unrolled Self-Supervised Learning}
\label{sec:ss}
Considering the limited generalization capabilities of supervised learning methods for the task at hand, our proposed approach substitutes supervised training with a self-supervised learning scheme. Instead of training a model that maps $\mathbf{y}_i$ to $\mathbf{\hat{x}}_i$, we train a sequence-specific, model-based autoencoder, which maps $\mathbf{y}_i$ to itself. The input $\mathbf{y}_i$, interpolated to the output size and reshaped to a $N \times N$ image, is fed to an autoencoder, which is comprised of two parts: an encoder which encodes $\mathbf{y}_i$ to sparse code $\mathbf{\hat{x}}_i$ (i.e., approximated emitters' locations), and a decoder which decodes $\mathbf{\hat{x}}_i$ to an approximation of the input $\mathbf{\hat{y}}_i$. 

\begin{figure}[!t]
\begin{minipage}[b]{1.0\linewidth}
  \centering
  \centerline{\includegraphics[scale=0.5]{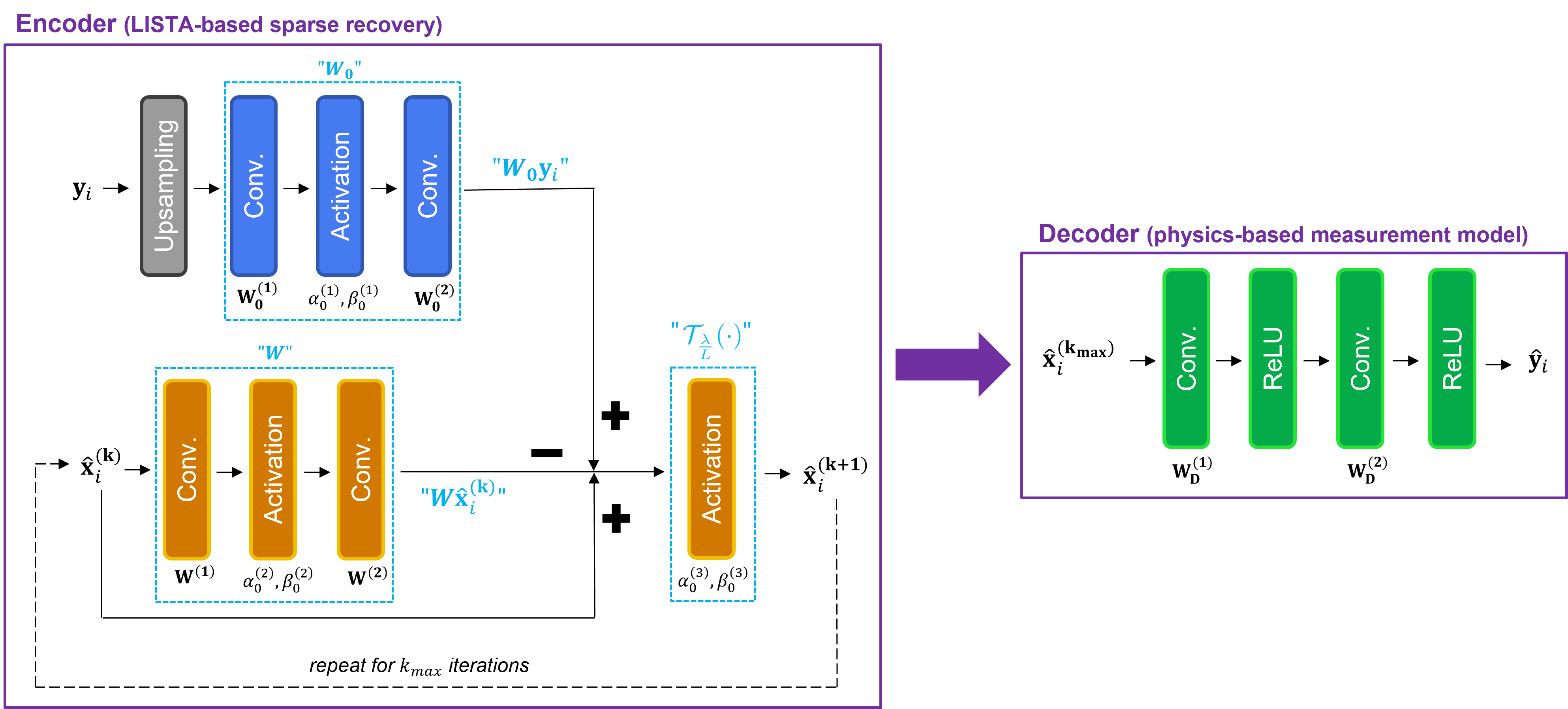}}
\end{minipage}
\caption{{\small Architecture of the model-based autoencoder. First, the input $\mathbf{y}_i$ is upsampled according to the given scale factor. Then, it is fed into a LISTA-based encoder, which performs sparse recovery to approximate $\mathbf{x}_i$. It is comprised of convolutional layers with trainable weights $\{\mathbf{W_{0}}^{(i)}, \mathbf{W}^{(i)}\}_{i=1,2}$, and activation layers $S^{+}_{\alpha_0, \beta_0}(\cdot)$ with trainable parameters $\{\alpha_{0}^{(i)}, \beta_{0}^{(i)}\}_{i=1...3}$. The first three layers (shown in blue) produce an initial approximation $\mathbf{\hat{x}^{(0)}}_i$, which is then iteratively modified via the latter four layers (shown in orange). After $k_{max}$ iterations, the approximated sparse code $\mathbf{\hat{x}^{(k_{max})}}_i$ is fed through a decoder that mimics the physical measurement process, via convolution with learned filters. It has two blocks of convolutional layers with trainable weights $\{\mathbf{W_{D}}^{(i)}\}_{i=1,2}$, followed by ReLU activations. The decoder outputs $\mathbf{\hat{y}}_i$, which is an approximation of the input image.}}
\label{fig:lista}
\end{figure}

The architecture of both the encoder and the decoder is model-based, as illustrated in Figure \ref{fig:lista}. First, the LISTA-based encoder, whose layers resemble a single ISTA iteration. The thresholding step of the ISTA algorithm is replaced with the smooth activation function $S^{+}_{\alpha_0, \beta_0}(\cdot)$, with trainable parameters $\alpha_0, \beta_0$, as defined in (\ref{eqn:activ}) and (\ref{eqn:activ_params}). Each matrix-multiplication operation in (\ref{eqn:ISTA_step}) is replaced by a convolution operation with a learned kernel. Furthermore, to reduce the number of parameters in the model and avoid over-parameterized kernels, while still retaining a large enough receptive field, we decomposed each convolution operation into three layers: convolution, activation and another convolution, where each convolution kernel is significantly smaller. Dardikman et al. used kernels of size $29 \times 29$ and $25 \times 25$ in LISTA-based LSPARCOM \cite{LSPARCOM}; here, we use kernels of size $15 \times 15$. To mimic several iterations of ISTA, we repeat the same layers, with the same trainable weights and parameters, for a set number of iterations $k_{max}$. This keeps the number of model parameters to a minimum and enables efficient training and inference.

The decoder is a simple convolutional model which mimics the actual, physical \say{decoding} process of the data. We assume an image formation model of the form $\mathbf{y}_i = \mathbf{Ax}_i$, where each column $j$ of $\mathbf{A}$ represents the PSF of the system for a point source at location $j$. Thus, the multiplication operation $\mathbf{Ax}_i$ is equivalent to convolving the high-resolution  $\mathbf{x}_i$ with the system's PSF, with stride of size $\frac{N}{M}$, creating a down-sampling effect. We consider this process as a single convolution operation. Similar to the encoder, we reduce the number of parameters in the model and avoid over-parameterized kernels by decomposing the convolution operation into three layers: convolution, activation and another convolution. This results in a shallow convolutional neural network, with two convolutional layers followed by ReLU activations. This allows for the decoder to emulate the physical image generation process, with its parameters optimized to map the sparse code generated by the encoder to the actual measurements.

The full forward pass of the model-based autoencoder is described in Algorithm \ref{alg:fp}. It takes as input the low-resolution measurement $\mathbf{y}_i$ and number of iterations $k_{max}$. The input is interpolated to match the resolution of  $\mathbf{x}_i$, and is then put through the encoder's initial convolutional and activation layers, with filters $\{\mathbf{W_{0}}^{(i)}\}_{i=1,2}$ and parameters $\alpha_{0}^{(1)}, \beta_{0}^{(1)}$, respectively. This is equivalent to the first matrix multiplication operation in (\ref{eqn:ISTA_step}), $\mathbf{W_{0}}\mathbf{y}_i$. The result of the first three layers is then fed through another activation layer, with parameters $\alpha_{0}^{(3)}, \beta_{0}^{(3)}$, which is equivalent to the positive soft thresholding operation in Algorithm \ref{alg:ISTA}. This results in an initial estimation of $\mathbf{x}_i$, denoted as $\mathbf{\hat{x}^{(0)}}_i$. The approximation of $\mathbf{x}_i$, denoted as $\mathbf{\hat{x}^{(k)}}_i$, is iteratively updated via the latter convolutional and activation layers, with filters $\{\mathbf{W}^{(i)}\}_{i=1,2}$ and parameters $\alpha_{0}^{(2)}, \beta_{0}^{(2)}$, respectively. This update is equivalent to the second matrix multiplication operation in (\ref{eqn:ISTA_step}), $(\mathbf{I-W})\mathbf{\hat{x}^{(k)}}_i$. Following the same arithmetic logic of the original ISTA update step, the outputs of the first three layers and the latter three layers are combined and fed through the final activation layer, with parameters $\alpha_{0}^{(3)}, \beta_{0}^{(3)}$. This yields the updated approximation of $\mathbf{x}_i$, denoted as $\mathbf{\hat{x}}^{(k+1)}_i$. Upon reaching $k_{max}$ iterations, The approximated $\mathbf{\hat{x}}^{(k_{max})}_i$ is fed through the decoder's two blocks of convolution with filters $\{\mathbf{W_{D}}^{(i)}\}_{i=1,2}$, followed by ReLU activations. The decoder's output, denoted as $\mathbf{\hat{y}}_i$, is the approximation of the low-resolution input measurement.

\begin{algorithm}[H]
\setstretch{1}
\caption{Forward Pass}
\label{alg:fp}
\begin{algorithmic}[1]
\Require{$\mathbf{y}_i$, number of iterations $k_{max}$ }
\Ensure{$\mathbf{\hat{y}}_i$}
\State $\mathbf{y}_i = $Upsample$(\mathbf{y}_i)$
\State $\mathbf{\hat{x}^{(0)}}_i =  S^{+}_{\alpha_{0}^{(3)}, \beta_{0}^{(3)}}(\mathbf{W_{0}}^{(2)} * (S^{+}_{\alpha_{0}^{(1)}, \beta_{0}^{(1)}}(\mathbf{W_{0}}^{(1)} * \mathbf{y}_i)))$
\While{$k<k_{max}$}
\State $\mathbf{\hat{x}}^{(k+1)}_i = S^{+}_{\alpha_{0}^{(3)}, \beta_{0}^{(3)}}(\mathbf{\hat{x}}^{(k)}_i - \mathbf{W}^{(2)} * (S^{+}_{\alpha_{0}^{(2}, \beta_{0}^{(2)}}(\mathbf{W}^{(1)} *\mathbf{\hat{x}}^{(k)}_i)) + \mathbf{W_{0}}^{(2)} * (S^{+}_{\alpha_{0}^{(1)}, \beta_{0}^{(1)}}(\mathbf{W_{0}}^{(1)} * \mathbf{y}_i)))$
\State $k\leftarrow k+1$
\EndWhile
\State $\mathbf{\hat{y}}_i = \textrm{ReLU}(\mathbf{W}_{D}^{(2)} * \textrm{ReLU}(\mathbf{W}_{D}^{(1)} * \mathbf{\hat{x}}^{(k_{max})}_i))$
\end{algorithmic}
\end{algorithm}

Training of the autoencoder is fully self-supervised, since it maps its input to itself. Therefore, our training set consists only of $\{\mathbf{y}_i\}_{i=1,...,T}$. The weights are optimized using $L_1$ loss with the ADAM optimizer \cite{adam}, given $\beta_1 = 0.9, \beta_2 = 0.999$ and an initial learning rate of $1e^{-3}$, using the PyTorch library. The trainable weights of the convolutional layers were initialized according to their respective tasks: $\{\mathbf{W_{D}}^{(i)},\mathbf{W_{0}}^{(i)},\mathbf{W}^{(i)}\}_{i=1,2}$, which imitate multiplication by $\mathbf{A}, \mathbf{A}^T$ and $\mathbf{A}^T\mathbf{A}$ respectively, were initialized as a $15\times15$ Gaussian filter with $\sigma = 1$. The activation layers are initialized $\alpha_0=0.95, \hspace{0.25em} \beta_0=8$ for all layers.


\section{Results}
In this section, we compare the reconstruction quality achieved using Self-STORM to that achieved by other methods for SMLM reconstruction: iterative algorithm, SPARCOM \cite{SPARCOM} and its unrolled, supervised version, LSPARCOM \cite{LSPARCOM}; prominent learning-based localization methods for high emitter-density frame sequences, Deep-STORM \cite{deepSTORM} and DECODE \cite{decode}; and a generic self-supervised technique, based on ZSSR \cite{zssr}. All algorithms tested for comparison take high emitter-density frame sequences as input, and thus allow similarly high temporal resolution. Therefore, the comparison focuses on the quality of the reconstruction based on the exact same input frames. When possible, we also compare the results to the ground truth localization. We further compare runtimes for all algorithms; all timings were conducted while running on a Nvidia Tesla V100 32GB GPU. For all figures, the colormap is such that white corresponds to the highest value, then yellow, red and black. For the ground-truth, Self-STORM, ZSSR, Deep-STORM and DECODE reconstructions, the value obtained corresponds to the integrated emitter intensity. For the SPARCOM and LSPARCOM reconstructions, the value obtained corresponds to the variance of the emitter. Since the values per pixel is usually not of interest, but rather only the support of the image which indicates the location of the emitters, the maximal and minimal values mapped to the edges of the colormap at each image were chosen to obtain optimal visibility.

\subsection{Pre-processing, Training and Inference}

\subsubsection{Self-STORM}
As explained in section 4, Self-STORM utilizes an autoencoder to infer high-resolution localizations $\mathbf{\hat{x}}_i$ from low-resolution measurements $\mathbf{y}_i$, for $i = 1, ..., T$. During training, the autoencoder maps $\mathbf{y}_i$ to itself; therefore, our training set consists only of $\{\mathbf{y}_i\}_{i=1,...,T}$. During inference, we ignore the decoder and only use the encoder to approximate $\mathbf{\hat{x}}_i$ from $\mathbf{y}_i$. Finally, the support of the sum over $\{\mathbf{\hat{x}}_i\}_{i=i...T}$ gives us the super-resolved image of localized molecules. The chosen scale factor for the super-resolved images is $\frac{N}{M} = 4$. The maximal number of unrolled LISTA iterations in the encoder was empirically optimized during both training and inference. The best results were obtained with $k_{max} = 1$ during training and $k_{max} = 2$ during inference (for all datasets). Training of the model was stopped after 2-8 epochs (optimized per dataset to achieve the best possible result).

Pre-processing of the data was included a designated normalization scheme, designed to address the fact that emitter concentrations and their intensities often vary significantly across different parts of the frame sequence. Standard image normalization techniques lead to unequal representation of different regions in the FOV, to the point where a significant amount of the emitters are inseparable from background noise. Therefore, we used a normalization scheme tailored for this problem: first, we label the pixels of each frame according to their value, where the top percentiles of each frame are labeled as non-background, and the rest of the pixels are labeled as background. The exact percentile chosen as threshold was optimized to obtain the best possible result per dataset. Second, we detect the \say{centroids} in each frame: non-background pixels whose value is higher than all their neighbors. Then, we set a bounding box around each centroid according to the non-background pixels surrounding it. Finally, the pixels within each bounding box are normalized according to their mean and standard deviation.

\subsubsection{LSPARCOM}
In LSPARCOM \cite{LSPARCOM}, the super-resolved localization image is generated by inputting a single image, which is constructed by calculating the temporal variance of all the low-resolution inputs. We used a pretrained version of LSPARCOM, as released by their authors \cite{gilli}. LSPARCOM was trained on a publicly-available simulated dataset of 12,000, $64 \times 64$ frames, composing a single FoV with an underlying structure of biological microtubules \cite{smlm_rev}; the ground truth positions can be seen in Figure \ref{fig:BTHD}(a). To increase emitter density per frame, random combinations of 40 frames from the original sequence were summed together, generating a new sequence of 360 high density frames. The training set itself contains 10,000, $16 \times 16$ patch-stacks, randomly extracted from the high density frame sequence. Pre-processing of the data, both prior to training and testing, includes normalizing the movie intensity to have a maximal value of 256, and removing the temporal median of the movie from each frame.

\subsubsection{Deep-STORM}
In Deep-STORM \cite{deepSTORM}, each low-resolution input is processed independently, and the final high-resolution reconstruction is achieved via summation, similar to our proposed method. The training set contains 10,000 random $26 \times 26$ patches, extracted from 20, $64 \times 64$ simulated images containing randomly positioned emitters. These images were generated via the ThunderSTORM ImageJ plugin \cite{ThunderSTORM}, ideally with the exact same imaging parameters as the test set. These parameters include the camera base level, photo-electrons per A/D count, PSF, emitter FWHM range, emitter intensity range and mean photon background. Pre-processing in Deep-STORM includes resizing the input frame to the desired dimensions of the output (determining the final grid size), projecting each frame to the range $[0 \ 1]$, and normalizing each frame by removing the mean value of the training dataset and dividing by its standard deviation, as specified in the code released by the authors \cite{zcdl4mic}.

\subsubsection{DECODE}
The DECODE model \cite{decode} takes each low-resolution frame as input, along with its temporally-neighboring frames as temporal context. For each frame, it predicts multiple channels: the first two channels indicate the probability for the presence on an emitter in each pixel, as well as its intensity. The next three channels describe the sub-pixel coordinates of the emitter (with respect to the center of the pixel). An additional channel predicts the background intensity in each pixel. The super-resolved localization image is generated by aggregating the results from all low-resolution frames on a high-resolution grid, according to the desired scale factor (which was set to $\frac{N}{M} = 4$ in our case). We used a pretrained version of DECODE, as released by their authors \cite{decode}. The model was trained on 20000 simulated images containing randomly positioned emitters, generated by a custom simulation provided by the authors \cite{decode}. Similar to Deep-STORM, the imaging parameters of the simulation are also ideally set to be the same as those of the test set. There is no pre-processing on the input data.

\subsubsection{ZSSR}
Self-supervised approaches for single image super-resolution, such as ZSSR \cite{zssr}, essentially follow the same scheme presented above for Self-STORM. The only major difference is in the architecture of the learned model. In Self-STORM, our architecture is based on deep unrolling of an iterative algorithm. This is not the case for other methods (ZSSR included). which use a generic convolutional neural network, therefore not exploiting any prior domain knowledge on the problem at hand and designing the model's architecture accordingly. Thus, to demonstrate the significance of using a model-based architecture as opposed to a generic one, we train an autoencoder similar to the one described in Section \ref{sec:ss}, but with a simple convolutional neural network as its encoder, instead of the LISTA-based model. In this case, the encoder is composed of six consecutive convolutional layers, each followed by ReLU activation. All other parameters, pre-processing, training and inference schemes are the same as previously described for Self-STORM.

\subsubsection{SPARCOM}
In SPARCOM \cite{SPARCOM}, similar to LSPARCOM, the super-resolved localization image is generated by inputting a single image, which is constructed by calculating the temporal variance of all the low-resolution inputs. We used the classical version of SPARCOM, executed over 100 iterations, without using a weighing matrix or applying the sparsity prior in another transformation domain \cite{SPARCOM_MATH}, to keep the basic LISTA algorithm similar to the one used for unrolling. To achieve optimal performance, we used fast ISTA (FISTA) \cite{FISTA} and the Fourier-domain implementation of the original algorithm \cite{SPARCOM_MATH}. The regularization factor $\lambda$ was hand-picked and fine-tuned to fit each tested dataset. SPARCOM also requires explicit prior knowledge of the PSF: for the simulated data, the PSF used for generating the data was accurately given; for the experimental data we used the same PSF as the simulated data (having no better choice, since the actual PSF is unknown). Similar to LSPARCOM, pre-processing in SPARCOM includes normalizing the movie intensity to have a maximal value of 256, and removing the temporal median of the movie from each frame.

\subsection{Simulation Results}
In this subsection, we evaluate the methods using simulated realistic data, where we have precise knowledge of the ground truth localizations. Having access to the ground truth enables measuring the quality of the results using the SNR metric, which was also previously used in the SMLM challenge \cite{smlm_rev}. The SNR is defined as follows:
\begin{equation}
\begin{gathered}
\label{eqn:snr}
    \text{SNR} \equiv 10\log_{10} \frac{\|\mathbf{X}_{GT}\|^2}{\|\mathbf{X}_{GT} - \mathbf{\hat{X}}\|^2},
\end{gathered}
\end{equation}
where $\mathbf{X}_{GT}$ is the ground-truth binary image of molecule locations, and $\mathbf{\hat{X}}$ is the approximated super-resolved image of localized molecules. Since this metric compares locations (i.e., binary images), all output images were binarized prior to calculating the SNR; the thresholding step was optimized to obtain the best possible score per each method and dataset.

\begin{figure*}[t!]
\begin{minipage}[b]{1.0\linewidth}
  \centering
  \centerline{\includegraphics[scale=0.375]{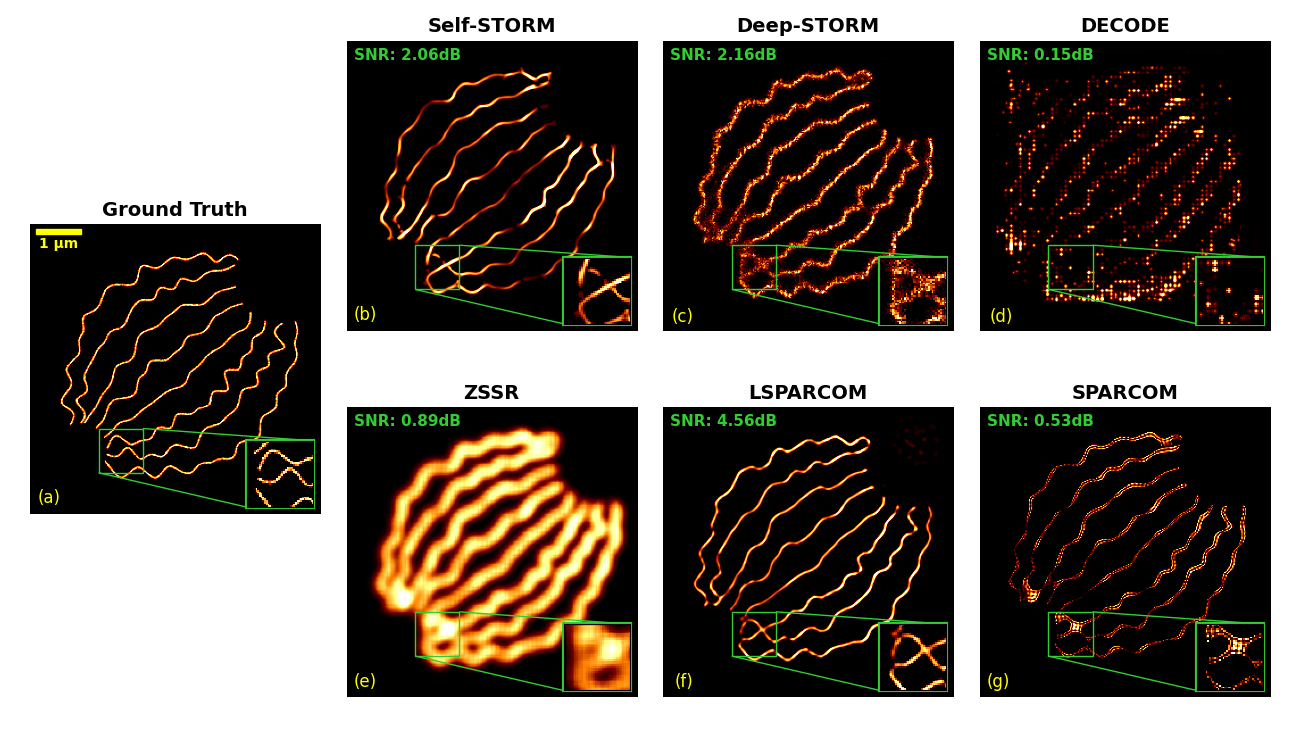}}
\end{minipage}
\caption{\small Super-resolved reconstruction of a simulated microtubules dataset \cite{smlm_rev}, composed of 361 high density frames. (a) Ground truth. (b) Self-STORM reconstruction. (c) Deep-STORM reconstruction. (d) DECODE reconstruction. (e) ZSSR reconstruction. (f) LSPARCOM reconstruction. (g) SPARCOM reconstruction executed over 100 iterations with $\lambda = 0.0105$. SNR is shown in the upper-left corner of each reconstructed image. It is evident that LSPARCOM gives the best results, as it was trained on data with identical ground-truth structure. Self-STORM and Deep-STORM achieve similar visual reconstruction quality, while all other methods yield far less accurate results.}
\label{fig:BTHD}
\end{figure*}

Figure \ref{fig:BTHD} shows the results for a dataset of 361 high density frames, whose structure is identical to the one used for creating the training set of LSPARCOM, with a slightly different set of imaging parameters (simulating imaging with a different microscope). As can be seen, Self-STORM yields results which are second only to LSPARCOM (which was trained on data with the exact same structure), in terms of both SNR and visual resemblance of ground-truth structure. Deep-STORM achieves slightly higher SNR than Self-STORM, but fails to properly separate close microtubules to the same degree as Self-STORM and LSPARCOM. Other methods are not nearly as accurate, as the ZSSR-based reconstruction, which utilizes generic self-supervision without incorporating any prior knowledge to the learned model, merely smears the input image without pinpointing the locations of the molecules. The DECODE and SPARCOM reconstructions are both grainy and inaccurate, underperforming all other methods. In terms of runtime, the \textit{training and inference} time of Self-STORM and ZSSR was 31.83 sec. By comparison, Deep-STORM reconstruction took 4.32 sec, the DECODE reconstruction took 15.54 sec, the LSPARCOM reconstruction took 2.73 sec, and the SPARCOM reconstruction took 10.41 sec for 361, $64 \times 64$ input frames.

\begin{figure*}[h!]
\begin{minipage}[b]{1.0\linewidth}
  \centering
  \centerline{\includegraphics[scale=0.375]{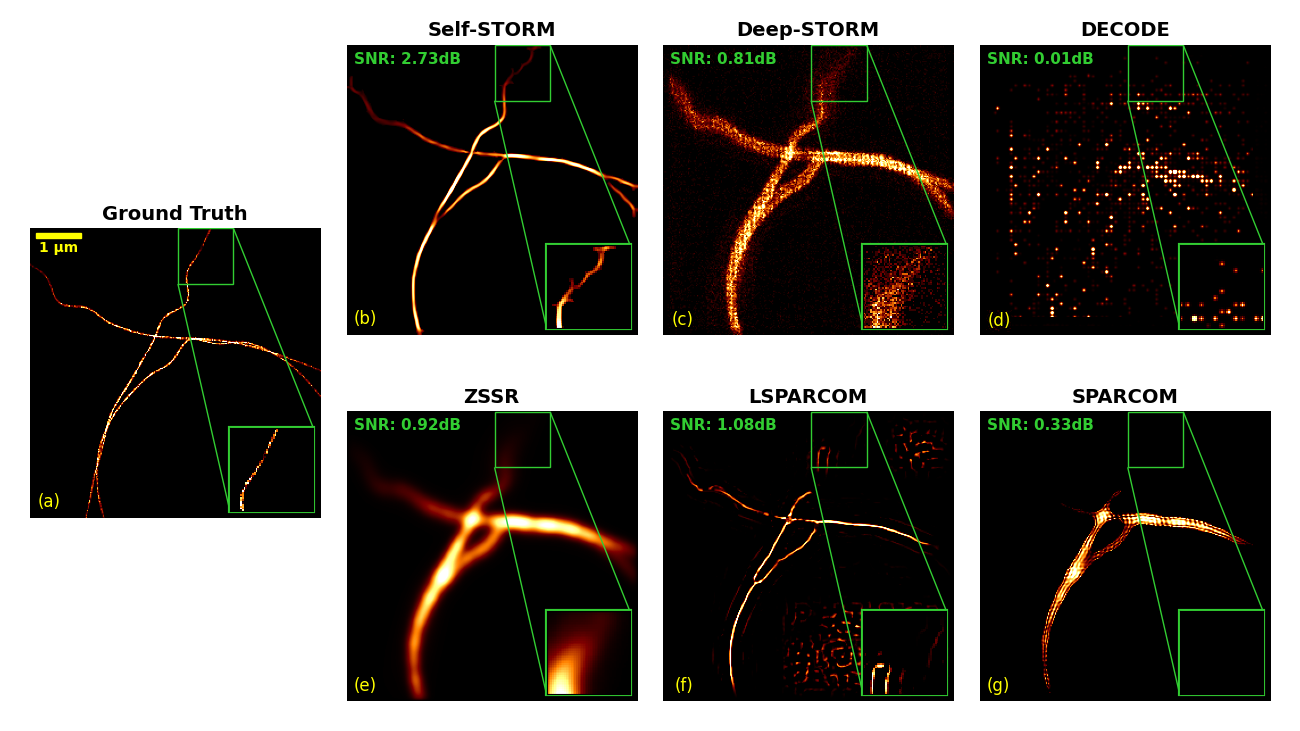}}
\end{minipage}
\caption{\small Super-resolved reconstruction of a simulated microtubules dataset \cite{smlm_rev}, composed of 2500 high density frames. (a) Ground truth. (b) Self-STORM reconstruction. (c) Deep-STORM reconstruction. (d) DECODE reconstruction. (e) ZSSR reconstruction. (f) LSPARCOM reconstruction. (g) SPARCOM reconstruction executed over 100 iterations with $\lambda = 0.003$. SNR is shown in the upper-left corner of each reconstructed image. In this case, Self-STORM provides an accurate reconstruction of the ground-truth, while all other methods fail to achieve similar reconstruction quality.}
\label{fig:MTO}
\end{figure*}

Figure \ref{fig:MTO} shows the results for a dataset of 2500 high density frames, whose structure and imaging parameters are different than any of the training sets used to train LSPARCOM, Deep-STORM or DECODE. In this case, Self-STORM is clearly the top performer, having the highest SNR (by a very large margin of over 1.5dB) and being the most visually similar to the ground-truth structure. In the enlarged region, it is evident that Self-STORM is very close to accurately localize the emitters in that region, where other methods fail to be anywhere close to the ground-truth positions. Deep-STORM, DECODE and LSPARCOM were all trained on substantially different data than this dataset, which led to their poor performance in this case. The ZSSR-based method produces a blurry, imprecise  reconstruction, as it did for the first dataset. The SPARCOM reconstruction is also very inaccurate and partial, compared to its higher-performing learning-based counterparts. In terms of runtime, the training and inference time of Self-STORM and ZSSR was 48.76 sec. By comparison, Deep-STORM reconstruction took 9.21 sec, the DECODE reconstruction took 27.31 sec, the LSPARCOM reconstruction took 3.11 sec, and the SPARCOM reconstruction took 12.56 sec for 2500, $64 \times 64$ input frames.

\begin{figure*}[h!]
\begin{minipage}[b]{1.0\linewidth}
  \centering
  \centerline{\includegraphics[scale=0.375]{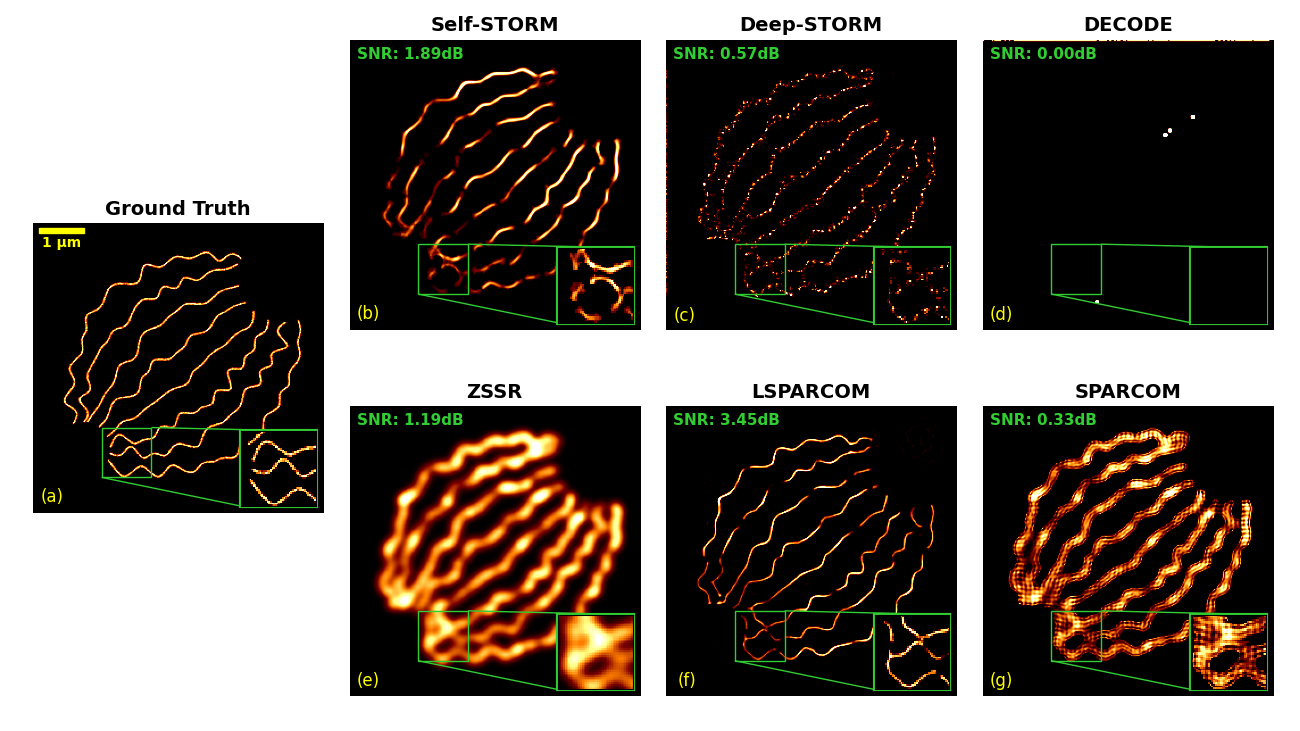}}
\end{minipage}
\caption{\small Super-resolved reconstruction of a simulated microtubules dataset \cite{smlm_rev}, composed of 36 ultra-high density frames, generated by summing every 10 consecutive frames of the original dataset. (a) Ground truth. (b) Self-STORM reconstruction. (c) Deep-STORM reconstruction. (d) DECODE reconstruction. (e) ZSSR reconstruction. (f) LSPARCOM reconstruction. (g) SPARCOM reconstruction executed over 100 iterations with $\lambda = 0.00105$. SNR is shown in the upper-left corner of each reconstructed image. Even with as few as 36 frames of ultra-high emitter density, Self-STORM manages to successfully reconstruct the underlying structure of the image without using any prior training data. Similar to Figure \ref{fig:BTHD}, it outperforms every other method besides LSPARCOM, which was on data with identical ground-truth structure.}
\label{fig:BTHD_X10}
\end{figure*}

Figure \ref{fig:BTHD_X10} presents the results for a 10x denser dataset than the one shown in Figure \ref{fig:BTHD}, obtained by summing every 10 consecutive frames in the original dataset, to a total of 36 highly dense input frames. This extremely dense input leads to substantial degradation in reconstruction quality for most methods, as the increase in density has made the reconstructions appear more fragmented. Self-STORM and LSPARCOM are the only two methods that manage to produce adequate reconstructions compared to their results on the original dataset. In terms of SNR, Self-STORM seems to maintain its accuracy despite the increased emitter density, with only a slight decrease of 0.17dB; LSPARCOM, on the other hand, has a major decrease of 1.11dB. In terms of runtime, the training and inference time of Self-STORM and ZSSR was 5.09 sec. By comparison, Deep-STORM reconstruction took 2.12 sec, the DECODE reconstruction took 3.71 sec, the LSPARCOM reconstruction took 3.07 sec, and the SPARCOM reconstruction took 9.85 sec for 36, $64 \times 64$ input frames. It is evident that the decrease in number of input frames has shortened the inference time of the methods operating on a frame-by-frame basis. On the other hand, the running time of SPARCOM and LSPARCOM has remained almost the same, as the number of input frames does not affect the runtime.

\begin{figure*}[h!]
\begin{minipage}[b]{1.0\linewidth}
  \centering
  \centerline{\includegraphics[scale=0.375]{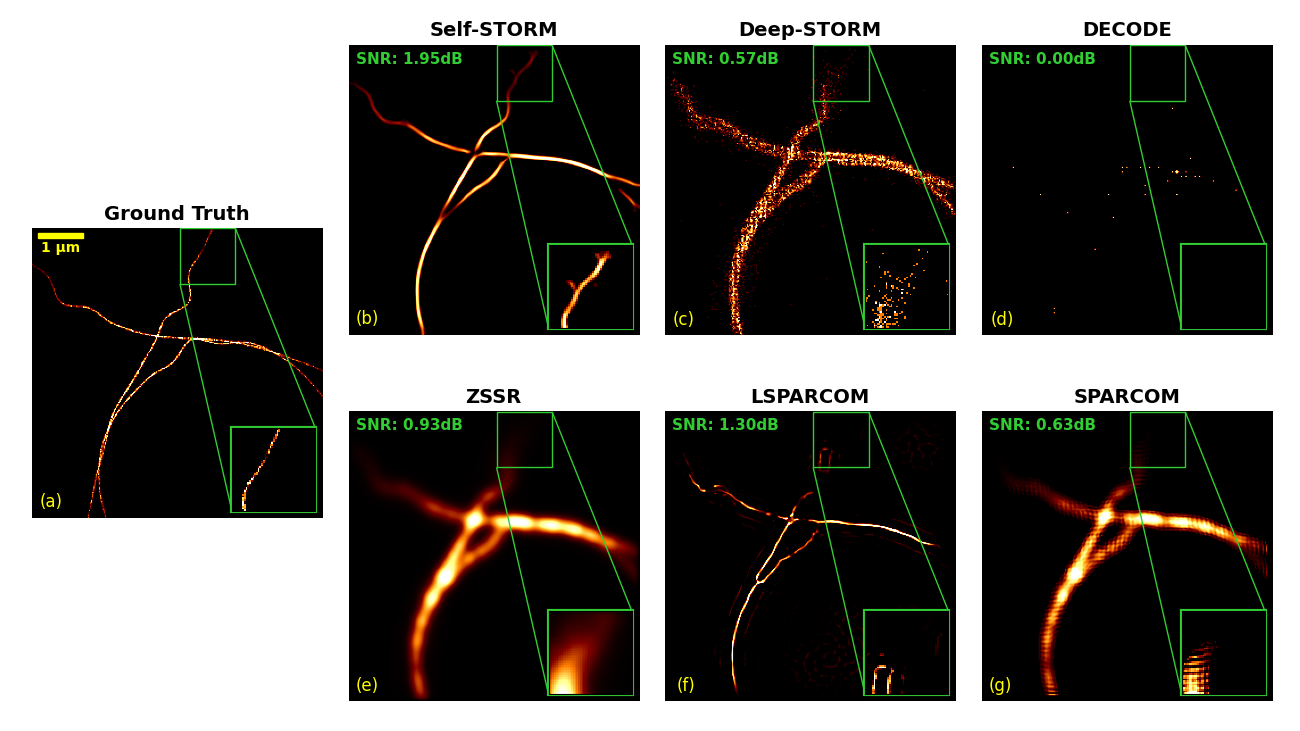}}
\end{minipage}
\caption{\small Super-resolved reconstruction of a simulated microtubules dataset \cite{smlm_rev}, composed of 250 ultra-high density frames, generated by summing every 10 consecutive frames of the original dataset. (a) Ground truth. (b) Self-STORM reconstruction. (c) Deep-STORM reconstruction. (d) DECODE reconstruction. (e) ZSSR reconstruction. (f) LSPARCOM reconstruction. (g) SPARCOM reconstruction executed over 100 iterations with $\lambda = 0.0003$. SNR is shown in the upper-left corner of each reconstructed image. Depsite the ultra-high density of emitters, Self-STORM provides a fairly accurate reconstruction of the ground-truth, suffering from slight degradation in quality compared to the results in Figure \ref{fig:MTO}. All other methods perform far worse, both in terms of SNR and visual resemblance of the ground-truth image.}
\label{fig:MTO_X10}
\end{figure*}

Figure \ref{fig:MTO_X10} presents the results for a 10x denser dataset than the one shown in Figure \ref{fig:MTO}, obtained by summing every 10 consecutive frames in the original dataset, yielding an overall of 250 highly dense input frames. Similar to the results on the previous dataset, it is evident that the great increase in emitter density leads to degraded reconstruction quality for all methods. Self-STORM is still the only method that succeeds in reconstructing the full underlying structure of the data, as it has for the original dataset. The Self-STORM reconstruction is the most similar the the ground truth image, compared to every other reconstruction. In the enlarged region of the image, Self-STORM demonstrates a remarkable ability to localize emitters very close to their ground-truth positions, outperforming other methods which struggle to achieve proximity to the ground-truth positions. However, in terms of SNR, Self-STORM is considerably less accurate, with a decrease of 0.78dB compared to its result on the original dataset.  In terms of runtime, the training and inference time of Self-STORM and ZSSR was 14.78 sec. By comparison, Deep-STORM reconstruction took 5.98 sec, the DECODE reconstruction took 12.47 sec, the LSPARCOM reconstruction took 4.02 sec, and the SPARCOM reconstruction took 8.97 sec for 250, $64 \times 64$ input frames. As in the case of the previous dataset, the reduction in total number of input frames has accelerated the inference time for methods operating frame-by-frame. The runtime SPARCOM and LSPARCOM runtime has remained consistent, unaffected by the total number of input frames.

In summary, the simulations shown illustrate that Self-STORM yields a robust and precise reconstruction, on par with or surpassing other methods, especially when dealing with data that is considerably different from the training data of these methods. It does so without any external training samples to learn from, nor any prior knowledge regarding the PSF or imaging parameters of the system. Its total runtime (training and inference) is comparable to the inference time of tested methods, and it is able to produce fairly accurate reconstruction with as few as 36 input frames with very high emitter density.

\subsection{Experimental Results}
In this subsection, we compare all methods using publicly-available experimental data. Since there is no ground truth image provided for these sequences of high density input frames, we evaluate the results qualitatively. We focus on the ability of each method to reconstruct fine details in the image and their visual resemblance to the input frames (i.e., the diffraction-limited image).

\begin{figure*}[h!]
\begin{minipage}[b]{1.0\linewidth}
  \centering
  \centerline{\includegraphics[scale=0.375]{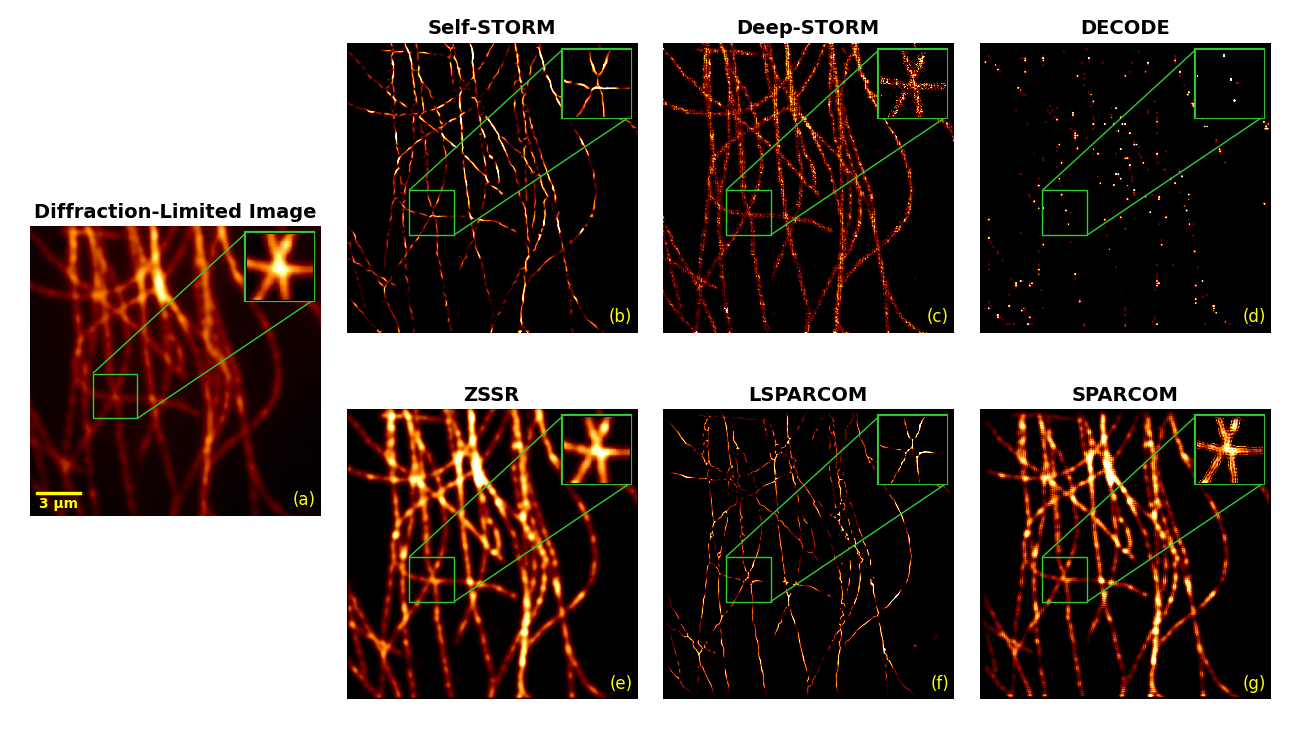}}
\end{minipage}
\caption{\small Super-resolved reconstruction of an experimental tubulins dataset \cite{smlm_rev}, composed of 500 high density frames. (a) Diffraction-limited image. (b) Self-STORM reconstruction. (c) Deep-STORM reconstruction. (d) DECODE reconstruction. (e) ZSSR reconstruction. (f) LSPARCOM reconstruction. (g) SPARCOM reconstruction, executed over 100 iterations with $\lambda = 0.0005$. Self-STORM achieves the most similar visual reconstruction of the diffration-limited image. Other methods yield subpar results, with visible fragmentation and/or artifacts in the reconstructed images.}
\label{fig:microtubules}
\end{figure*}

Figure \ref{fig:microtubules} shows the results for an experimental tubulins dataset of 500 high density frames, whose structure and imaging parameters are unknown and different than any of the training sets used to train LSPARCOM, Deep-STORM or DECODE. Self-STORM seems to obtain the best results, in terms of visual similarity to the diffraction-limited image, and lack of any fragmentation/artifacts in the reconstructed image. LSPARCOM achieves similar results but seems to be more fragmented, while Deep-STORM and SPARCOM exhibit less accurate reconstructions. DECODE completely fails to reconstruct the underlying structure of data, resulting in a highly fragmented image. The ZSSR-based reconstruction produces a blurry reconstruction that is not any better than the diffraction-limited image, as it did for the simulated datasets. In terms of runtime, the training and inference time of Self-STORM and ZSSR was 55.38 sec. By comparison, Deep-STORM reconstruction took 17.45 sec, the DECODE reconstruction took 34.61 sec, the LSPARCOM reconstruction took 7.21 sec, and the SPARCOM reconstruction took 31.15 sec for 500, $128 \times 128$ input frames.

\begin{figure*}[h!]
\begin{minipage}[b]{1.0\linewidth}
  \centering
  \centerline{\includegraphics[scale=0.375]{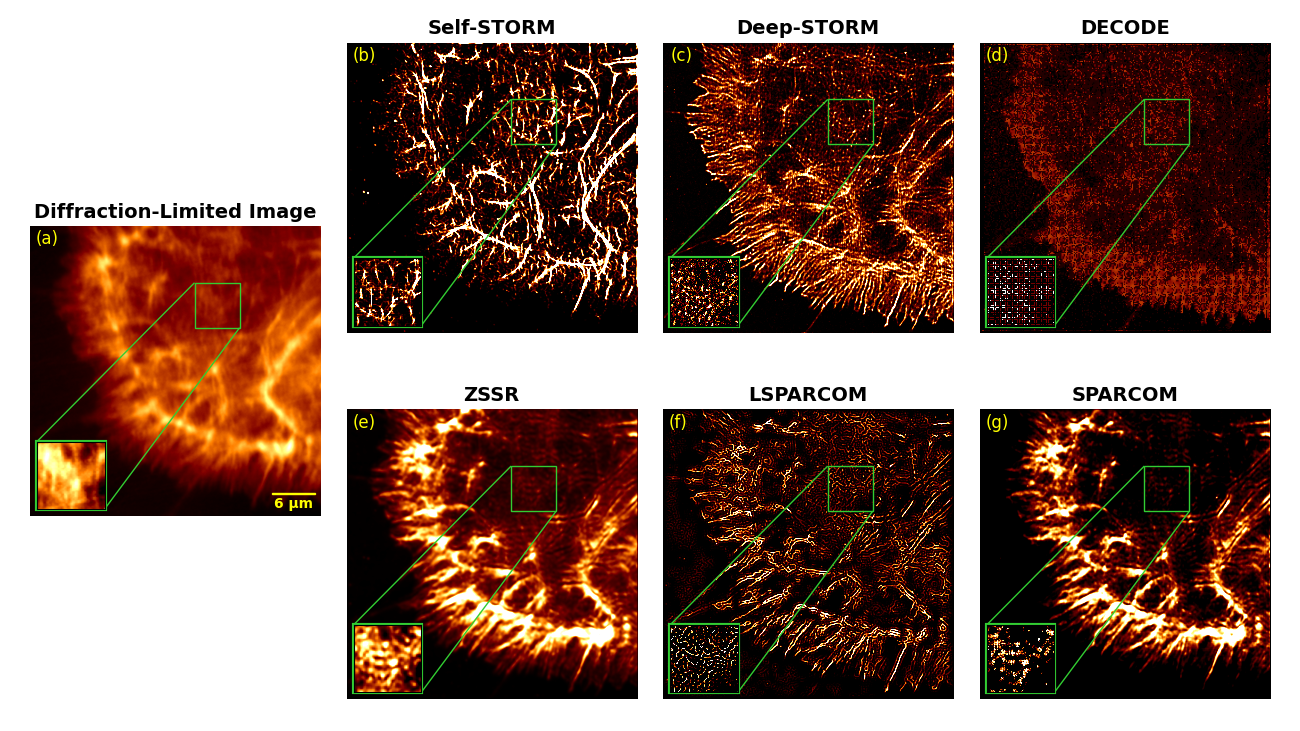}}
\end{minipage}
\caption{\small Super-resolved reconstruction of an experimental dataset capturing a glial cell in a culture of rat hippocampal neurons \cite{gila}, composed of 599 ultra-high density frames. (a) Diffraction-limited image. (b) Self-STORM reconstruction. (c) Deep-STORM reconstruction. (d) DECODE reconstruction. (e) ZSSR reconstruction. (f) LSPARCOM reconstruction. (g) SPARCOM reconstruction, executed over 100 iterations with $\lambda = 0.0005$. It is evident that Self-STORM succeeds in resolving the underlying structure of the glial cell up to very fine details, where other methods fail to do the same.}
\label{fig:Glia_actin}
\end{figure*}

Figure \ref{fig:Glia_actin} shows the results for an experimental tubulins dataset ocapturing a glial cell in a culture of rat hippocampal neurons \cite{gila}, composed of 599 ultra-high density frames. The structure and imaging parameters of this dataset are unknown and different than any of the training sets used to train LSPARCOM, Deep-STORM or DECODE. The original dataset contained 14975 frames; it was temporally binned in groups of 25 consecutive time frames in order to simulate ultra-high density of emitters. The Self-STORM reconstruction is most visually similar to the diffraction-limited image, and also captures fine details of the glial cell structure, as shown in the enlarged region. LSPARCOM and Deep-STORM also produce reconstructions that are fairly similar to the input image, but fail to accurately resolve the fine structure of the cell. ZSSR and SPARCOM both yield reconstructions that are rather blurry and partial that do not reveal the full, fine structure of the cell. DECODE yields a reconstruction that is notably grainy and imprecise, exhibiting the least resemblance to the diffraction-limited image. In terms of runtime, the training and inference time of Self-STORM and ZSSR was 118.97 sec. By comparison, Deep-STORM reconstruction took 31.15 sec, the DECODE reconstruction took 49.87 sec, the LSPARCOM reconstruction took 19.61 sec, and the SPARCOM reconstruction took 78.85 sec for 599, $256 \times 256$ input frames.

In conclusion of the experimental results section, Self-STORM excels in reconstructing experimental data with high emitter density, outperforming every other tested method while maintaining a comparable total runtime. The lack of dedicated training (for LSPARCOM, Deep-STORM and DECODE), model-based architecture (for ZSSR), or prior knowledge of the PSF (for SPARCOM) is evident in the inferior performance of other methods. In contrast, Self-STORM produces good reconstructions without requiring any of the above.


\section{Discussion and Conclusions}
The core concept of deep unrolled self-supervised learning is to use the algorithmic framework dictated by the setting of the problem to gain interpretability and domain knowledge, while inferring optimal parameters from input data only. This approach disrupts the trade-off between reliance on input data characteristics and dependence on prior knowledge and hyperparameter adjustment. Classical, non-learning-based algorithms such as SPARCOM have low dependency on the type of input data (as long as it fits the sparse prior), but are highly dependent on accurate knowledge of the PSF and additional hyperparameter tuning. On the other hand, standard deep-learning based methods such as DECODE, Deep-STORM and LSPARCOM do not require hyperparameter refinement (LSPARCOM also does not require exact knowledge of imaging parameters), but have very strong dependencies on the type of input data. Unlike other data-driven methods, Self-STORM is optimized per input, making it suitable for data of any type of structure and imaging parameters more than others. Figure \ref{fig:MTO} is a good example of this robustness: Self-STORM provided a full, precise reconstruction of the ground-truth image, while other tested methods produced results of very low quality, some of which are  completely irrelevant (due to mismatch between the training and test set). This ability is crucial, since it is not always feasible to generate a specific training dataset for each and every imaging setting.

It is also important to note the importance of incorporating both model-based learning (via deep algorithm unrolling) and self-supervision into the pipeline, as opposed to utilizing only one of these key concepts. This is demonstrated by the poor performance of the ZSSR-based method, whose generic model architecture results in reconstructions that are smeared, blurry versions of the diffraction-limited input. On the other hand, the use of model-based learning alone (i.e., in a supervised fashion), as in LSPARCOM, suffers from the aforementioned problem of performance degradation for test data that is considerably different than its training dataset. The proposed approach also leads to a compact model, allowing for fast training and inference, during test time, which is on the same time scale as the inference time of other methods. Moreover, as demonstrated in Figures \ref{fig:BTHD_X10} and \ref{fig:MTO_X10}, excellent reconstruction can even be obtained by Self-STORM using a relatively small number (few hundreds to a few dozens) of high emitter-density input frames, which allows for high temporal resolution and reduces the total runtime, making the difference between Self-STORM and other methods almost negligible. This is essential in enabling live-cell imaging via SMLM, where dynamic interactions on short time scales are of interest.

Self-STORM also has several limitations. First, as evident from the results in Figures \ref{fig:BTHD} and \ref{fig:BTHD_X10}, Self-STORM does not reach the same level of accuracy when compared to supervised learning methods, on data that is similar to their training sets. This is of course very reasonable, given that supervised models were fitted to yield optimal results for a specific type of input data, using ground-truth labels. Given a relevant training set, this gap in performance may be addressed by pre-training some of the weights of the model (the LISTA-based encoder, for example) via supervised learning. Then, in order to achieve optimal results per each specific input, the per-trained model may be fine-tuned using the self-supervised learning scheme. Second, the reconstruction of Self-STORM is sometimes fragmented, especially for very dense input sequences (see Figure \ref{fig:BTHD_X10}); this can be overcome by using a smoother regularizer (e.g. total variation) for deriving the iterative scheme used for unrolling, as shown in \cite{SPARCOM_MATH}. The spatial sparsity prior is also less suitable for samples of discrete nature, like receptors spread on the surface of a cell. This type of sample is more suitable for reconstruction considering sparsity in the wavelet domain \cite{SPARCOM_MATH}. Similarly to the fragmentation problem, this can be addressed by unrolling the appropriate sparse recovery algorithm. Finally, Self-STORM requires empirical selection of two parameters: the number of unrolled iterations $k_{max}$ (once during training and once during inference), and the number of training epochs. The number of unrolled iterations is a relatively easy choice, since it is very clear that the model performs well when trained with a single unrolled iteration. This is likely due to the optimal gradient computation and back-propagation it allows for, as gradients are back-propagated through each iteration. Therefore, increasing the number of unrolled iterations can result in very small gradients due to the repeated multiplication of gradients in each iteration. Interestingly, during inference, results improve when using two iterations instead of just one, aligning with the classical notion where more iterations lead to better convergence. Yet, adding additional iterations beyond two had negligible impact on the results. However, these choices might not be optimal if a different sparse recovery algorithm is used for unrolling. The second parameter is subject to more change, as different input datasets require a different number of training epochs to achieve optimal performance. To avoid manual stopping of the training process, an automatic stopping criteria may be used, similar to other self-supervised learning schemes \cite{dip, zssr, DualBSR, smore, sisr_rev}.

To conclude, Self-STORM offers a new method for SMLM data analysis, and sparse recovery in general, via deep unrolled self-supervised learning. Tested on a variety of datasets with various imaging parameters and geometries, Self-STORM has proven its ability to perform high-resolution localization for any given data. By comparison, other techniques fail to achieve similar performance on data that is significantly different from their external, labeled training sets. Thus, given its robust capabilities, Self-STORM has great potential for localization of biological structures, potentially replacing its counterparts for super-resolved imaging at the nanometer scale. On a more general scope, the combination of self-supervision and model-based learning may be advantageous for any sparse recovery problem: given a mathematical model that describes the measurement process and that the information to be recovered is sparse, one may construct the appropriate model-based encoder and decoder. The resulting pipeline may result in enhanced performance and specifically robustness to data heterogeneity, without the need for external training samples of any sort.


\bibliographystyle{IEEEbib}
\bibliography{main}


\end{document}